\documentclass{article}
\usepackage{spconf,amsmath,graphicx,amsfonts}
\usepackage{algorithm,algorithmic}
\usepackage[hidelinks]{hyperref}

\newcommand{\Mod}{\text{mod}\,}

\title{Binary sequence set optimization for CDMA applications via mixed-integer quadratic programming}

\name{Alan Yang\textsuperscript{1} \quad Tara Mina\textsuperscript{1} \quad Grace Gao\textsuperscript{2}}
\address{
\textsuperscript{1}Department of Electrical Engineering, Stanford University\\
\textsuperscript{2}Department of Aeronautics and Astronautics, Stanford University \\ \emph{\{yalan, tymina, gracegao\}@stanford.edu}}
\newcommand\blfootnote[1]{%
  \begingroup
  \renewcommand\thefootnote{}\footnote{#1}%
  \addtocounter{footnote}{-1}%
  \endgroup
}
\usepackage{fancyhdr}
\begin{document}
\ninept
\maketitle
\begin{abstract}
Finding sets of binary sequences with low auto- and cross-correlation properties is a hard combinatorial optimization problem with numerous applications, including multiple-input-multiple-output (MIMO) radar and global navigation satellite systems (GNSS). The sum of squared correlations, sometimes referred to as the integrated sidelobe level (ISL), is a quartic function in the variables and is a commonly-used metric of sequence set quality. In this paper, we show that the ISL minimization problem may be formulated as a mixed-integer quadratic program (MIQP). We then present a block coordinate descent (BCD) algorithm that iteratively optimizes over subsets of variables. 
The subset optimization subproblems are also MIQPs which may be handled more efficiently using specialized solvers than using exhaustive search; this allows us to perform BCD over larger variable subsets than previously possible.
Our approach was used to find sets of four binary sequences of lengths up to 1023 with better ISL performance than Gold codes and sequence sets found using existing BCD methods.
\blfootnote{The views expressed are those of the authors and do not
reflect the official guidance or position of the United States Government,
the Department of Defense or of the United States Air Force.
Statement from DoD: The appearance of external hyperlinks does not constitute endorsement by the United States Department of Defense (DoD) of the linked websites, or the information, products, or services contained therein.  The DoD does not exercise any editorial, security, or other control over the information you may find at these locations.}
\end{abstract}
\section{Introduction}
\label{sec:intro}

Code division multiple access (CDMA) is a multiple access technique that allows multiple signals to occupy a common communication channel~\cite{HeLS2012, VelazquezV2016}. 
In CDMA, all transmitters broadcast at the same carrier frequency, but each transmitter modulates the data signal with a unique and pre-determined (typically binary-valued) spreading code sequence.
A wide variety of applications currently utilize CDMA, including wireless and cellular network communications~\cite{abu2007introduction}, multi-input multi-output~(MIMO) radar systems~\cite{AlaeeMN2019,LinLi2020}, and Global Navigation Satellite Systems~(GNSS)~\cite{morton2021pnt}.

The choice of spreading code sequences directly influences the performance of the CDMA system.
In particular, to extract the signal from a particular transmitter, the received signal is correlated with a local replica of the spreading code.
A strong correlation peak indicates the presence of the corresponding transmitter's signal, and by aligning the local replica to the transmitter's received signal, the corresponding transmitted data can be recovered.
In order to reduce interference from different transmitters during signal extraction, it is desirable for the transmitters' spreading codes to have low cross-correlations with one another.
In addition, it is desirable for the spreading codes to have low autocorrelation at non-zero relative temporal shifts. If a spreading code correlates strongly with a temporally shifted version of itself, receivers are susceptible to false temporal alignment as a result of multipath interference. 
Therefore, it is desirable for the set of spreading codes to have both good cross-correlation and autocorrelation properties.

Algorithmically generable, or algebraic spreading codes such as Gold codes~\cite{Gold1967}, Kasami codes~\cite{Kasami1966}, and $m$-sequences~\cite{Proakis2001} have been widely used in wireless communications applications including MIMO radar and GNSS. The main advantage of algorithmically generable codes is that they can be produced on-the-fly (for example, using shift registers) and do not need to be stored in memory. However, the aforementioned codes are limited to only certain lengths of $2^n-1$ where $n$ is a natural number, and their autocorrelation and cross-correlation performance, as measured by the integrated sidelobe level~(ISL) metric~\cite{HeLS2012, AlaeeMN2019}, is suboptimal, in particular when the number of code sequences is much smaller than the sequence length. 

Over the years, decreasing memory storage costs have relaxed the requirement that spreading codes be algorithmically generable. It has become practical to store entire sets of spreading codes in memory, and there has been increasing interest in optimizing sequence sets of specific sizes and lengths to fit specific applications~\cite{VelazquezV2016}. 
In this work we consider binary spreading codes, that is, we search for sets of binary sequences with desirable correlation properties.

Block coordinate descent (BCD) is an approach that has been successfully applied to various binary sequence set design problems~\cite{AlaeeMN2019, CuiYFHL2017, LinLi2020, HuangL2020}.
BCD iteratively solves subproblems in which subsets of $N$ binary variables are optimized with the others held fixed; the subproblems are frequently solved via exhaustive search. 
However, since the subproblem search space grows exponentially with $N$, there is a practical limit on how large $N$ can be, if the subproblems are solved by exhaustive search.

In this paper, we introduce an approach that allows us to perform BCD-based binary sequence set optimization with larger variable subset sizes that would otherwise not be possible with exhaustive search. 
We first show that the cross-correlation function may be expressed as a linear function of the variables by adding auxiliary variables and linear inequality constraints. This allows us to formulate the binary sequence set design problem as a mixed-integer quadratic program (MIQP). In our approach, we minimize a version of the integrated sidelobe level~(ISL) objective, which is a common metric for evaluating binary sequence sets~\cite{HeLS2012, AlaeeMN2019, HuangL2020, alaee2018binary}. The ISL consists of a sum of squared correlation values, and is typically expressed as a non-convex quartic function.


In our formulation, the BCD subproblems are also MIQPs. This structure allows specialized solvers based on branch-and-bound, such as Gurobi~\cite{Gurobi}, to quickly solve subproblems involving larger variable subset sizes than previously possible with exhaustive search.
We used our approach to find sets of four binary sequences of lengths up to 1023 that have better ISL than possible using existing BCD methods, by optimizing over subsets of 20 variables at a time. 
Our approach is closely related to the class of methods proposed by Yuan et al., which combines block coordinate descent with exact combinatorial search, for solving discrete optimization problems \cite{YuanSZ2017}.


The rest of the paper is organized as follows. Sections~\ref{sec:prior_work}~and~\ref{sec:notation} review prior work and develop the binary sequence set optimization problem, respectively. Our representation of cross-correlation and proposed formulation for the ISL minimization problem are presented in Section~\ref{sec:linearization}, BCD is discussed in Section \ref{sec:bcd}, and experimental results are presented in Section~\ref{sec:experiments}. 

\section{Prior Work}
\label{sec:prior_work}

Continuous optimization techniques, such as penalty methods~\cite{YuCYLK2020} and semidefinite relaxations~\cite{DeMaioDHZF2008} have been proposed to design sets of complex-valued, continuous-phase sequences with constant magnitude. Continuous-valued sequences need to be discretized in practice, and binary sequences are often preferred due to ease of implementation~\cite{AlaeeMN2019,CuiYFHL2017}. Since the discretization of continuous sequences has been found to give poor performance, BCD methods have been proposed to directly optimize binary sequence sets for various applications~\cite{AlaeeMN2019, CuiYFHL2017, LinLi2020, HuangL2020}. Our work enables BCD methods to optimize over larger variable subset sizes at a time, which can lead to improved performance. 

Both Bose~and~Soltanalian~\cite{BoseS2018} and Boukerma~et~al.~\cite{BoukermaRMA2021} construct new sequences and sequence sets by combining pre-existing binary sequences with desirable correlation properties, such as Gold codes or optimized sequence sets; those approaches may be directly combined with BCD methods. Population-based methods, such as genetic algorithms~\cite{LiuSW2022} and natural evolution strategies~\cite{MinaG2022} have also been developed, although these methods do not consider the structure in the objective, instead treating it as a general nonlinear function.

\section{Definitions and Notation}\label{sec:notation}



We represent a set of~$K$ length-$L$ binary sequences using a matrix~$X\in\{\pm 1\}^{L\times K}$. Each column of $X$ represents one of the~$K$ sequences in the set. In what follows, we use the notation~$X_i$ to denote the $i$\textsuperscript{th}~column of~$X$, and we refer to ``columns,'' ``sequences,'' and ``codes'' interchangeably. Furthermore, we use the notation $X_{m,i}$ to denote the $m$\textsuperscript{th}~entry of the $i$\textsuperscript{th}~sequence, or the $m$\textsuperscript{th} row of the $i$\textsuperscript{th} column. For convenience, the indices are defined to start at zero.

We denote the cross-correlation between columns $i$ and $j$ of $X$ at shift $k$ by
\begin{equation}\label{eq:cross_correlation_set}
    \left(X_i \star X_j\right)_k = 
	\sum_{m=0}^{L-1} X_{m,i}\cdot X_{(m+k)_{\Mod L}, j}.
\end{equation}
We refer to the cross-correlation of column $i$ with itself as the autocorrelation of column $i$.
In this work, we exclusively consider the periodic autocorrelation and cross-correlation functions, although the extension to the aperiodic case is straightforward.

An ideal sequence set $X$ has correlation values $\left(X_i \star X_j\right)_k$ that are simultaneously close to zero for all values of $i$, $j$, and $k$, except for autocorrelations of shift $k=0$. From \eqref{eq:cross_correlation_set}, we see that $\left(X_i \star X_i\right)_0 = L$ for any $X_i\in\{\pm 1\}^L$.
In this work, we minimize a sum of squared correlation values
\begin{equation}\label{eq:isl}
	f(X) = \sum_{i=0}^{K-1} \sum_{j=i}^{K-1} \sum_{k=0}^{L-1} \left(X_i \star X_j\right)_k^2 \mathbf{1}_{\left\{i\ne j\,\text{or}\, k\ne 0\right\}},
\end{equation}
where $\mathbf{1}_{\{\cdot\}}$ denotes the indicator function. We refer to this objective function as the integrated sidelobe level~(ISL), due to its close similarity with functions of the same name defined in prior works~\cite{HeLS2012,AlaeeMN2019,HuangL2020}.
The indicator function ensures that zero-shift autocorrelations $\left(X_i \star X_i\right)_0$ are not included in the objective.
The sequence set design problem is to find an $X\in \{\pm 1\}^{L\times K}$ such that $f(X)$ is small. Note that $f$ is a quartic, non-convex function of $X$, and the problem is NP-hard.



\section{Reformulating ISL minimization as a MIQP}\label{sec:linearization}

In Subsection~\ref{subsec:xor_linearization}, we show that, by adding auxiliary variables, the cross-correlation \eqref{eq:cross_correlation_set} may be replaced by a linear function of the variables, subject to linear inequality constraints. In Subsection~\ref{subsec:isl}, this fact is used to reformulate the ISL minimization problem as a mixed-integer quadratic program (MIQP). In Subsection~\ref{subsec:bnb}, we introduce branch-and-bound methods, which may in principle be used to solve MIQPs.


\subsection{A linearization of cross-correlation}\label{subsec:xor_linearization}
Consider two binary variables~$a$~and~$b$, which may take values in~$\{\pm 1\}$. The product~$a\cdot b$ is bilinear in the variables, and is therefore not convex in~$a$~and~$b$. However, since $a$~and~$b$ are binary we may represent their product using an auxiliary variable~$z$ if we impose a set of linking constraints:
\begin{subequations}
\begin{align}
	z &\le b - a + 1, \label{eq:xor_constraint_start}\\
	z &\le a - b + 1, \\
	z &\ge -1 - a - b, \\
	z &\ge -1 + a + b. \label{eq:xor_constraint_end}
\end{align}
\end{subequations}
One may verify that $z=a\cdot b$ for each of the four combinations of values for $a$ and $b$, if and only if~$z$ satisfies~\eqref{eq:xor_constraint_start}~--~\eqref{eq:xor_constraint_end}~\cite{GloverW1974}.

We now introduce a set of auxiliary variables
$Z_{m,k}^{i,j}$ that satisfy the linking constraints
\begin{subequations}
\begin{align*}
	Z_{m,k}^{i,j} &\le X_{(m+k)_{\Mod L}, j} -  X_{m,i} + 1, \\
	Z_{m,k}^{i,j} &\le X_{m,i} - X_{(m+k)_{\Mod L}, j} + 1, \\
	Z_{m,k}^{i,j} &\ge -1 - X_{m,i} - X_{(m+k)_{\Mod L}, j}, \\
	Z_{m,k}^{i,j} &\ge -1 + X_{m,i}+ X_{(m+k)_{\Mod L}, j},
\end{align*}
\end{subequations}
which ensure that $Z_{m,k}^{i,j} = X_{m,i}\cdot X_{(m+k)_{\mod L}, j}$.
Subject to those constraints, the cross-correlation~\eqref{eq:cross_correlation_set} may therefore be written as
\begin{equation}\label{eq:cross_correlation_set_z}
    \left(X_i \star X_j\right)_k = \sum_{m=0}^{L-1} Z_{m,k}^{i,j},
\end{equation}
which is linear in the (auxiliary) variables.



\subsection{ISL minimization}\label{subsec:isl}

Inserting~\eqref{eq:cross_correlation_set_z}~into the ISL objective~\eqref{eq:isl} leads to the problem
\begin{subequations}\label{eq:isl_miqp}
	\begin{align}
		\mbox{minimize} &\quad
                \sum_{i=0}^{K-1} \sum_{j=i}^{K-1} \sum_{k=0}^{L-1}
                \left(\sum_{m=0}^{L-1} Z_{m,k}^{i,j}\right)^2
                \mathbf{1}_{\left\{i\ne j\,\text{or}\, k\ne 0\right\}}\\
		\mbox{subject to} 
		&\quad X \in\{\pm 1\}^{L\times K}, \label{eq:isl_miqp_con_binary} \\
		&\quad Z_{m,k}^{i,j} \le X_{(m+k)_{\Mod L}, j} -  X_{m,i} + 1, \label{eq:isl_miqp_con_start}\\
		&\quad Z_{m,k}^{i,j} \le X_{m,i} - X_{(m+k)_{\Mod L}, j} + 1, \\
		&\quad Z_{m,k}^{i,j} \ge -1 - X_{m,i} - X_{(m+k)_{\Mod L}, j}, \\
		&\quad Z_{m,k}^{i,j} \ge -1 + X_{m,i}+ X_{(m+k)_{\Mod L}, j},  \label{eq:isl_miqp_con_end} \\
		&\quad \text{for all}\,\, i,j,m,k. \nonumber
	\end{align}
\end{subequations}
This is a MIQP, since it involves the minimization of a convex quadratic function, subject to linear inequality constraints and binary constraints on $X$. The auxiliary variables $Z_{m,k}^{i,j}$ do not have explicit integer constraints, but are binary-valued in the feasible set.
Note that when the binary constraints \eqref{eq:isl_miqp_con_binary} are relaxed, the problem \eqref{eq:isl_miqp}~becomes a convex quadratic program (QP), which may be efficiently solved to give a lower bound on the optimal objective value~\cite{BoydV2004}. Adding additional linear inequality constraints or performing partial minimization over some of the variables with the others held fixed also lead to MIQPs, for which lower bounds are also available by solving QPs.


\subsection{Extension to general convex objectives}\label{subsec:convex_extension}

In Subsection~\ref{subsec:xor_linearization}, we showed that the correlation terms $(X_i\star X_j)_k$ terms may be replaced by linear functions of the variables.
Using that approach, we may replace the ISL objective $f(X)$ given in ~\eqref{eq:isl} with any other objective function $g(X)$ that is also convex in the correlation values~$\left(X_i\star X_j\right)_k$. By the affine pre-composition rule, $g$ will also be convex with respect to $X$.

For example, we may replace the square terms $\left(X_i\star X_j\right)_k^2$ in \eqref{eq:isl} with another convex function of $\left(X_i\star X_j\right)_k$, such as $\left|\left(X_i\star X_j\right)_k\right|$. Another example is the function $g(X) = \max_{i,j,k} \left|\left(X_i \star X_j\right)_k\right|$, which is referred to as the peak sidelobe level (PSL)~\cite{HeLS2012,AlaeeMN2019}.



\subsection{Branch-and-Bound}\label{subsec:bnb}

Branch-and-bound algorithms are commonly used for solving MIQPs~\cite{LawlerW1966}. During optimization, they maintain lower and upper bounds on the optimal objective value and return a solution that is provably optimal, up to a specified tolerance level. The bounds are used to rule out suboptimal regions in the search space and potentially solve the problem to optimality faster than exhaustive search. In this subsection, we give a brief sketch of the intuition behind the method; for more details see, for example, the references \cite{LawlerW1966, ConfortiCZ2014}.

As mentioned in Subsection~\ref{subsec:isl}, lower bounds may be obtained by relaxing integer constraints and solving the resulting QPs. Suppose that the relaxed solution has a variable $X_{i,j}$ that violates the integer constraints. If no such variable exists, then the relaxed solution is optimal. We proceed by choosing $X_{i,j}$ to be a branching variable, and form two new subproblems: one with $X_{i,j} = -1$, and another with $X_{i,j} = 1$. If we can solve the two subproblems to optimality, then the better of the two resulting solutions will be optimal for the original problem. In this way, we have replaced the original problem with two more tightly constrained (and therefore easier to solve) subproblems. This procedure may be repeated starting from each subproblem to form a search tree, where each node corresponds to a subproblem associated with a different branching variable.

A exhaustive search traverses all $2^{L\times K}$ nodes in the search tree. The idea in branch-and-bound is to leverage the bounds to prune the search tree. For example, if the lower bound at a node has objective value larger than the best feasible solution found so far, the entire sub-tree starting from that node may be eliminated from the search tree.

Commercial~\cite{Gurobi} and open-source~\cite{Juniper} solvers implement many sophisticated techniques and heuristics to accelerate branch-and-bound, and have been used to solve many real-world problems to optimality with reasonable speed \cite{ConfortiCZ2014}. Although the time complexity of branch-and-bound is exponential in the worst case, we expect the tightness of the lower bound obtained by convex relaxation to be a strong indicator of its performance in practice.


\subsection{Challenges with branch-and-bound}\label{subsec:bnb_issues}

Directly solving the MIQP~\eqref{eq:isl_miqp} using a commerical branch-and-bound solver such as Gurobi~\cite{Gurobi} is only practical for small problem instances.
The first challenge is symmetry~\cite{Margot2010}; permuting the columns of $X$ or circularly shifting the entries of a given column does not change the objective value. Second, note that if we relax the binary constraint~\eqref{eq:isl_miqp_con_binary}, the resulting QP gives a trivial lower bound of zero. Due to the aforementioned issues, it may take a very large number of branching steps to arrive at a subproblem with a useful lower bound obtained by convex relaxation. Moreover, the time needed to perform each branching step may become prohibitive, since the number of variables, linear inequality constraints, and terms in the objective function grows as $O(L^2K^2)$.

In the following section, we present a block coordinate descent algorithm that iteratively optimizes over subsets of the binary variables, while keeping the rest of the sequence set fixed. Indeed, the two aforementioned issues are in part circumvented if we settle for solving~\eqref{eq:isl_miqp} only over a subset of the variables. Fixing some of the variables to constant values can break symmetries and lead to more useful lower bounds for branch-and-bound.

\begin{algorithm}[t]
\caption{(Block) coordinate descent for sequence set design}
\begin{algorithmic}[1]
\STATE Choose an initial $X^{(0)}\in\{\pm 1\}^{L\times K}$, possibly at random
\STATE Initialize row index $i=0$ and column index $j=0$
\STATE Initialize $t=0$
\REPEAT
    \STATE $t\gets t+1$
    \STATE $S = \{(i, j)\}$
    \IF {\texttt{do\_BCD}}
        \STATE Choose another column $j'\ne j$ at random
        \STATE Choose indices $S'=\{(u_n,v_n)\mid n=1,\ldots,N-1\}$ at random, where each $u_n\in\{0,\ldots,L\}$ and $v_n\in\{j,j'\}$
        \STATE $S\gets S\cup S'$
    \ENDIF
    \STATE Set $X^{(t)}$ to be the solution of the minimization
    \begin{equation*}
        \begin{array}{ll}
            \mbox{minimize} & f(X) \\
            \mbox{subject to}
            & X\in\{\pm 1\}^{L\times K} \\
            & X_{i',j'} = X^{(t-1)}_{i',j'}, \,\, \forall (i',j')\not\in S
        \end{array}
    \end{equation*}
    \vspace{-3mm}
    \IF {$f(X^{(t)})$ has not improved in $LK$ steps}
        \STATE \textbf{break}
    \ELSIF {$f(X^{(t)})$ has not improved in $L$ steps}
        \STATE $j\gets (j + 1)_{\Mod K}$
    \ENDIF
    \STATE $i\gets (i+1)_{\Mod L}$
\UNTIL{convergence}
\STATE \textbf{return} $X$
\end{algorithmic}
\label{alg:bcd_bnb}
\end{algorithm}

\section{Block Coordinate Descent}\label{sec:bcd}

Block coordinate descent~(BCD) repeatedly solves the optimization problem~\eqref{eq:isl_miqp} over only a subset of the variables at a time, while keeping the others fixed. In each iteration, we choose a subset of variable indices~$S$
and solve~\eqref{eq:isl_miqp} with $X_{i,j}$ held fixed if $(i,j)\notin S$. Note that BCD is a descent method, since the objective value 

The BiST coordinate descent algorithm~\cite{AlaeeMN2019} optimizes a single entry~$X_{i,j}$ at a time, with the others held fixed. The row~$i$ is incremented at every iteration, and the column~$j$ is incremented when the column~$j$ reaches a local optimum, that is, the objective cannot be improved by changing any single row in the column. Algorithm~\ref{alg:bcd_bnb} illustrates the BiST algorithm, when $\texttt{do\_BCD}$ is false. 
When $\texttt{do\_BCD}$ is set to be true in Algorithm~\ref{alg:bcd_bnb}, we extend BiST to the BCD case.

In BCD, we optimize over a total of $N>1$ indices, including $(i,j)$. The additional indices are randomly selected from two columns $j$ and~$j'\ne j$, where $j'$ is also randomly chosen.
We solve the BCD subproblem in line 12 of Algorithm~\ref{alg:bcd_bnb} by instead solving \eqref{eq:isl_miqp}, using a branch-and-bound method. When the subset size is small, exhaustive search may be used instead; Cui et al. considered four variables at a time~\cite{CuiYFHL2017}. 

\subsection{Choosing the BCD subset size}\label{subsec:bcd_subset_size}

\begin{figure}[htb]
\begin{minipage}[b]{1.0\linewidth}
  \centering
  \centerline{\includegraphics[width=0.83\textwidth]{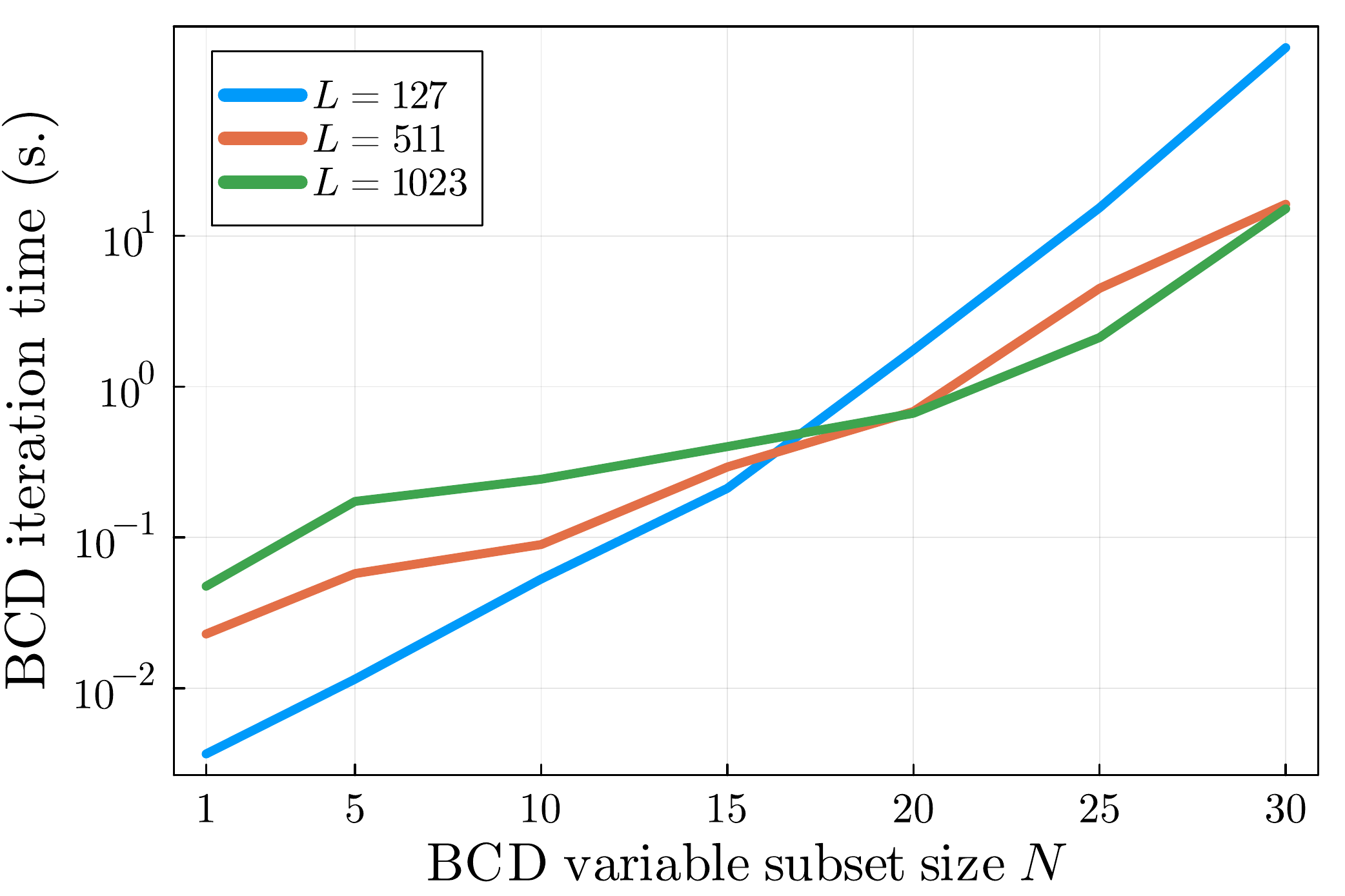}}
  \caption{Median BCD iteration time vs. variable subset size, for sets of $K=4$ sequences and varying sequence lengths $L$.}
\label{fig:bcd_timing}
\end{minipage}
\end{figure}

In general, increasing the variable subset size $N$ can lead to better performance, but lead to more expensive BCD steps.
Figure~\ref{fig:bcd_timing} compares the median time taken per iteration by BCD as the variable subset size $N$ is increased, for varying $L$ and $K=4$, over ten random variable subsets each. When the variable subset sizes are small ($N \le 15$), the iteration time increases with sequence length, as expected. However, for $N \ge 15$, the situation is reversed; choosing a variable subset size of 30 is more practical for $L=1023$ than it is for $L=127$. This may be explained by the issues discussed in Subsection~\ref{subsec:bnb_issues}. As the variable subset size approaches the sequence length, the quality of the relaxed lower bound is expected to decrease, which means that the number of branching steps is expected to increase.

The results in Figure~\ref{fig:bcd_timing}, as well as the results in the following section, were obtained using the Gurobi solver~\cite{Gurobi}, along with JuMP~\cite{DunningHL2017}, which is implemented using the Julia programming language.
Our code has been made publicly available\footnote{{\scriptsize\url{https://github.com/Stanford-NavLab/binary_seq_opt}}}.

\begin{table}[htb]
\centering
\begin{tabular}{l | cccc}
& \multicolumn{4}{c}{Sequence length $L$}\\[2pt]
& $63$ & $127$ & $511$ & $1023$ \\[2pt]
\hline
Gold & 27,506 & 123,538 & 2,053,810 & 8,784,498 \\
BiST & 26,018 & 104,418 & 1,692,626 & 6,778,098 \\
BCD ($N=4$) & 25,826 & 104,418 & 1,692,626 & 6,778,098 \\
BCD ($N=20$) & \textbf{25,386} & \textbf{103,930} & \textbf{1,690,906} & \textbf{6,769,906}
\end{tabular}
\caption{ISL minimization performance comparison for $K=4$ sets of sequences with varying $L$. BiST \cite{alaee2018binary} is equivalent to BCD with $N=1$.}
\label{table:bcd_performance_k4}
\end{table}

\section{Performance Comparison}
\label{sec:experiments}

We compared the performance of BCD using different variable subset sizes $N$ with BiST~\cite{AlaeeMN2019} and Gold codes~\cite{Gold1967}. Note that BiST is equivalent to BCD with $N=1$. First, we ran BiST for $L=63, 127, 511,$ and $1023$ all with $K=4$ until convergence, in each case starting from ten different randomly generated initial codes. For each $L$, the ten BiST solutions were then used as initial conditions for BCD, which we tested with variable subset sizes $N=4$ and $N=20$. 

When $N=4$, the BCD subproblem was solved using exhaustive search, similar to the approach taken by Cui et al.~\cite{CuiYFHL2017}. The Gurobi optimizer~\cite{Gurobi} was used for the $N=20$ case. We also compared with Gold codes \cite{Gold1967}, which are used by the Global Positioning System (GPS). For each $L$, we sampled one million random subsets of $K=4$ Gold codes, and chose the subset with the best ISL.

Table~\ref{table:bcd_performance_k4} compares the performance of the Gold codes, BiST, and BCD methods. For BiST and the BCD methods, the table shows the best ISL achieved out of the ten runs. In each case, increasing the variable subset size lead to improved solutions. Since BiST cannot improve after the objective has not decreased for $LK$ steps and the BCD methods were initialized from the outputs of BiST, the results indicate that increasing $N$ can consistently improve the performance of BCD. For the BCD methods, it is possible that better solutions could have been found by running the methods for more iterations.





\section{Conclusion}
\label{sec:conclusion}

In this paper, we showed that the cross-correlation function may be expressed as a linear function of the variables, subject to linear inequality constraints. Using this approach, we formulated the binary sequence set optimization problem as a MIQP. Our formulation allowed us to perform BCD over larger variable subsets than previously possible by using an MIQP solver. Finally, we demonstrated that our approach outperforms Gold codes and existing BCD methods on several binary sequence set optimization problems, relevant to MIMO radar, GNSS, and other CDMA applications. Possible directions for future work include alternate variable subset selection schemes and convex objective functions other than ISL.


\section{Acknowledgments}
\label{sec:acknowledgments}
This material is based upon work supported by the Air Force Research Lab (AFRL) under grant number FA9453-20-1-0002.


\bibliographystyle{IEEEbib}
\bibliography{coderefs}

\end{document}